\documentclass{ifacconf}

\usepackage{graphicx}           
\usepackage{tikz}               
\usepackage{pgfplots}           
\usepackage{pgfplotstable}      
\usepackage{fontawesome5}       
\usepackage[utf8]{inputenc}     
\usepackage{comment}            
\usepackage{xcolor}             
\usepackage{amsfonts}           
\usepackage{amsmath}            
\usepackage{mathtools}          
\usepackage{amssymb}            
\usepackage{bm}                 
\usepackage{booktabs}           
\usepackage{tabularx}           
\usepackage{multirow}           
\usepackage{makecell}           
\usepackage{setspace}           
\usepackage{colortbl}           
\usepackage{times}              
\usepackage{natbib}             
\usepackage{enumitem}           
\usepackage{cases}              
\usepackage{etoolbox}           
\usepackage{balance}            
\usepackage[hyphens]{url}       
\usepackage{lipsum}             

\allowdisplaybreaks 

\usetikzlibrary{calc,arrows.meta,positioning,shapes.geometric,fit,external}
\usepgfplotslibrary{groupplots,statistics,external}
\pgfplotsset{compat=1.18}

\definecolor{ggrnn_color}{RGB}{115, 0, 140}      
\definecolor{gru_color}{RGB}{44, 160, 44}        
\definecolor{lstm_color}{RGB}{190, 20, 20}       
\definecolor{mlp_color}{RGB}{230, 115, 0}        
\definecolor{real_color}{RGB}{0, 0, 0}           
\definecolor{input_color}{RGB}{34, 139, 34}      
\definecolor{state_color}{RGB}{210, 65, 20}      
\definecolor{output_color}{RGB}{0, 0, 0}         

\renewcommand{\captionwidth}{\linewidth}

\newcommand{\paragraphfont}[1]{\textsc{#1}.} 

\newtheorem{remark}{Remark}
\newtheorem{proposition}{Proposition}

\newbool{arxiv}
\setbool{arxiv}{false}   

\newcommand{\arxiv}[1]{\ifbool{arxiv}{{\color{red}#1}}{}}

\newcommand{\R}{\mathbb{R}}
\newcommand{\N}{\mathbb{N}}
\newcommand{\Z}{\mathbb{Z}}
\newcommand{\Zpos}{\Z^+}
\newcommand{\iset}[2]{\mathbb{I}_{#1}^{#2}}
\newcommand{\trans}{\top}
\newcommand{\gatedom}{\mathcal{G}}
\newcommand{\statedom}{\mathcal{H}}

\DeclarePairedDelimiter\pr{\lparen}{\rparen}
\DeclarePairedDelimiter\ps{\lbrack}{\rbrack}
\DeclarePairedDelimiter\pc{\lbrace}{\rbrace}

\DeclarePairedDelimiterX\norm[1]{\lvert}{\rvert}{%
    \ifblank{#1}{\,\cdot\,}{#1}%
}
\DeclarePairedDelimiter\abs{\lvert}{\rvert}

\DeclarePairedDelimiterXPP\set[3]{}{\lbrace}{\rbrace}{%
    \ifblank{#2}{}{_{#2}}\ifblank{#3}{}{^{#3}}%
}{%
    #1%
}

\newcommand{\hiddenunits}{x}
\newcommand{\nx}{n_{\hiddenunits}}
\newcommand{\nuin}{n_u}
\newcommand{\ny}{n_y}
\newcommand{\ns}{n_s}
\newcommand{\nenc}{n_{\mathrm{enc}}}

\newcommand{\parameters}{\theta}
\newcommand{\bparameters}{\bm{\parameters}}
\newcommand{\bParameters}{\bm{\Theta}}
\newcommand{\ds}{\mathcal{D}}
\newcommand{\loss}[1]{\mathcal{L}_{\mathrm{#1}}}

\newcommand{\examplesymbol}{\diamond}

\newcommand{\settablestretch}{\setstretch{1.02}}
\newcommand{\settablefontsize}{\footnotesize}

\begin{document}
\begin{frontmatter}

\title{
Generalized Gated Recurrent Neural Networks\\
for Nonlinear System Identification
} 

\thanks[corresp]{Corresponding author: \texttt{stefano.decarli@unibg.it}. This work has been submitted to IFAC for possible publication.}

\author[unibg]{Stefano De Carli\thanksref{corresp}}
\author[unibg]{\quad Davide Previtali}
\author[unibg]{\quad Mirko Mazzoleni}
\author[unibg]{\quad Fabio Previdi}

\address[unibg]{Department of Management, Information and Production Engineering, University of Bergamo, Italy}
\begin{abstract}
    \label{s:abstract}
    Long Short-Term Memory (LSTM) and Gated Recurrent Unit (GRU) networks are the two most widely adopted recurrent neural networks for black-box nonlinear system identification, yet they are invariably treated as mutually exclusive structural choices.
    To move toward a general framework that inherits the properties of both networks, we introduce the Generalized Gated Recurrent Neural Network (GGRNN), a unified architecture that embeds the exact GRU and LSTM network dynamics through homotopies between their respective gating mechanisms.
    Each homotopy continuously deforms one mechanism into the other, with a vector of homotopy parameters blending the contribution of the GRU and LSTM counterparts.
    Since benchmark comparisons report no consistent winner between the two architectures, the gating mechanism best suited to a given system is unlikely to coincide with either endpoint.
    We thus let each recurrent unit learn its own position along these homotopies, obtaining gating behaviors that neither endpoint can express.
    We pair the GGRNN with a two-stage training methodology: in Stage~1 we co-train all shared and network-specific parameters in parallel at the GRU and LSTM endpoint architectures, while in Stage~2 we freeze the network parameters and optimize only the homotopy parameters.
    Experiments on the Wiener--Hammerstein and Nano--Drone benchmarks show improved simulation accuracy over the co-trained GRU and LSTM networks on most output groups, and over the reference black-box baselines in nearly all comparisons.
\end{abstract}


\end{frontmatter}

\section{Introduction}
\label{s:introduction}

Gated Recurrent Neural Networks (GRNNs) have become the reference neural network models for black-box nonlinear system identification~\citep{pillonettoDeepNetworksSystem2025}.
Practically all applications rely on two canonical architectures: the Long Short-Term Memory (LSTM) network~\citep{hochreiterLongShortTermMemory1997} and the Gated Recurrent Unit (GRU) network~\citep{choLearningPhraseRepresentations2014}.
Both networks emerged in machine learning as alternative solutions to the same vanishing and exploding gradient issue that afflicts vanilla RNNs on long sequences~\citep{hochreiterLongShortTermMemory1997}.
They share the same core mechanism, i.e., an additive retention term that keeps error gradients alive along shortcut paths, yet implement it through different structural choices, first compared by~\citet{chungEmpiricalEvaluationGated2014} in their empirical evaluation of gated recurrent architectures.

Despite this common origin, the literature treats the two networks as competing alternatives and offers no clear selection criterion.
A common rule of thumb in the machine learning literature, dating back to empirical comparisons on speech and signal modeling tasks~\citep{chungEmpiricalEvaluationGated2014}, pairs LSTM networks with long-term dependencies and GRU networks with ``shorter" dynamics.
In nonlinear system identification, this heuristic has never been validated: physical systems mix multiple timescales in the data, and benchmark comparisons report no consistent winner between the two models~\citep[e.g.,][]{champneysBaselineResultsSelected2024, schoukensBenchmarkingNonlinearSystem2026}.
The machine learning literature is just as inconclusive: \citet{chungEmpiricalEvaluationGated2014} reported mixed, dataset-dependent results, whereas \citet{greffLSTMSearchSpace2017} found that no architectural variant consistently outperforms the standard LSTM network.

Architectural studies have so far explored only combinatorial structural changes, adding or removing whole gating mechanisms, i.e., the learned elementwise scalings that modulate the flow of information through the network.
Ad hoc gate removals yield simplified variants, such as the minimal gated unit~\citep{zhouMinimalGatedUnit2016} and chaos-free networks~\citep{decarliChaosFreeNetworksAre2026}.
Exhaustive combinatorial searches test thousands of gate configurations; their largest instance~\citep{greffLSTMSearchSpace2017} identified the forget gate and the output activation as the most critical components of the LSTM network.
Yet, rather than eliminating gates from the LSTM network as the GRU network does, a continuous unification that combines their complementary mechanisms within the same layer is still missing.

To overcome this dichotomy, we introduce the \emph{Generalized Gated Recurrent Neural Network} (GGRNN), a unified architecture that embeds the exact GRU and LSTM state space equations as the two endpoints of four elementwise homotopies between their gating mechanisms.
Each homotopy is governed by a vector of homotopy parameters that blends the contribution of the GRU and LSTM counterparts of each gating mechanism.
We identify the GGRNN with a two-stage methodology.
In Stage~1, we co-train all shared and network-specific parameters in parallel at the exact GRU and LSTM endpoints of these homotopies, establishing a common latent state space representation.
In Stage~2, we freeze all network parameters and optimize only the homotopy parameters.

We evaluate the proposed model on public nonlinear identification benchmarks~\citep{schoukensBenchmarkingNonlinearSystem2026}.
Specifically, we test the GGRNN on the classic Wiener--Hammerstein benchmark~\citep{schoukensWienerHammersteinBenchmark2009} and on the challenging Nano--Drone benchmark~\citep{busettoNonlinearSystemIdentification2026}, two systems whose nonlinear dynamics evolve across multiple timescales.
The GGRNN improves simulation accuracy over the co-trained GRU and LSTM networks on the Wiener--Hammerstein system and on three of the four cumulative errors for the Nano--Drone benchmark, while learning different combinations of gating mechanisms across recurrent units.

\paragraphfont{Organization}
The remainder of this paper is organized as follows.
Section~\ref{s:preliminaries} formalizes the identification problem and reviews the LSTM and GRU state space models.
Section~\ref{s:ggrnn} introduces the GGRNN architecture and its exact endpoint reductions.
Section~\ref{s:training_methodology} presents the two-stage training methodology.
Section~\ref{s:results} reports the numerical experiments on the benchmarks.
Finally, Section~\ref{s:conclusion} summarizes the main findings.

\paragraphfont{Notation}
Let $\R$, $\R_{\geq 0}$, and $\R_{>0}$ denote the sets of real, nonnegative real, and strictly positive real numbers, $\N$ be the set of non-negative integers, $\Zpos \coloneq \N \setminus \pc{0}$, and $\iset{a}{b} \coloneq \set*{a, a+1, \dots, b}{}{}$ for $a \leq b$, with $a, b \in \N$.
All-zero and all-one vectors and matrices of appropriate dimensions are denoted by $\bm{0}$ and $\bm{1}$, respectively.
For $\bm{a}, \bm{b} \in \R^n$, $\bm{a} \circ \bm{b}$ is the Hadamard product and $\norm{\bm{a}}$ the Euclidean norm.
The sigmoid and hyperbolic tangent functions are $\sigma\pr{z} \coloneq \pr*{\bm{1} + e^{-z}}^{-1}$ and $\phi\pr{z} \coloneq \tanh\pr{z}$, respectively, for $z \in \R$, applied elementwise to vector and matrix arguments.
%
\section{Preliminaries}
\label{s:preliminaries}
%
This Section presents the problem setting and the background material.
We first formalize the system identification problem in Section~\ref{ss:problem_statement}, then review the two canonical gated architectures in Section~\ref{ss:gated_rnns}.
%
\subsection{Problem Statement}
\label{ss:problem_statement}
%
We consider the identification of a nonlinear discrete-time Multiple-Input Multiple-Output (MIMO) dynamical system with input $\bm{u}_k \in \R^{\nuin}$, $\nuin \in \Zpos$, and output $\bm{y}_k \in \R^{\ny}$, $\ny \in \Zpos$, from data.
We define the true system as:
\begin{equation}
    \label{eq:true_system}
    \begin{aligned}
        \bm{s}_{k+1} &= \bm{f}_{\mathrm{true}}\pr*{\bm{s}_k, \bm{u}_k},\\
        \bm{y}_k &= \bm{g}_{\mathrm{true}}\pr*{\bm{s}_k, \bm{u}_k} + \bm{v}_k, 
    \end{aligned}
\end{equation}
where $k \in \N$ is the discrete time step, $\bm{s}_k \in \R^{\ns}$ is the true state of the system with dimension $\ns \in \Zpos$, $\bm{f}_{\mathrm{true}}: \R^{\ns} \times \R^{\nuin} \to \R^{\ns}$ is the state transition function, $\bm{g}_{\mathrm{true}}: \R^{\ns} \times \R^{\nuin} \to \R^{\ny}$ is the output map, and $\bm{v}_k \in \R^{\ny}$ denotes the additive measurement noise at time $k$.
The objective is to learn the input--output behavior of~\eqref{eq:true_system} from a dataset $\ds = \pc*{(\bm{u}_k, \bm{y}_k)}_{k=0}^{N-1}$ of length $N \in \Zpos$ via a GRNN architecture.
%
\subsection{Gated Recurrent Neural Networks}
\label{ss:gated_rnns}
%
To formulate both canonical networks in a unified state space form~\citep{pillonettoDeepNetworksSystem2025}, let $n_x \in \Zpos$ denote the state dimension, $\bm{x}_k \in \R^{n_x}$ the cell state, and $\bm{h}_k \in \R^{n_x}$ the hidden state.
For each gating or candidate state equation $\examplesymbol \in \pc{f, i, r, o, c}$, let $W_\examplesymbol \in \R^{n_x \times \nuin}$, $R_\examplesymbol \in \R^{n_x \times n_x}$, and $\bm{b}_\examplesymbol \in \R^{n_x}$ denote the input weight matrix, recurrent weight matrix, and bias vector, collected as $\bparameters_\examplesymbol \coloneq \pc*{W_\examplesymbol, R_\examplesymbol, \bm{b}_\examplesymbol}$.
Gates take values in $\gatedom \coloneq \pr{0,1}^{n_x}$ and hidden states in $\statedom \coloneq \pr{-1,1}^{n_x}$, with closed counterparts $\overline{\gatedom} \coloneq \ps{0,1}^{n_x}$ and $\overline{\statedom} \coloneq \ps{-1,1}^{n_x}$.

The LSTM network proposed by~\citet{hochreiterLongShortTermMemory1997} decouples $\bm{x}_k \in \R^{\nx}$ from $\bm{h}_k \in \statedom$ using a forget gate $\bm{f}_k \in \gatedom$, an input gate $\check{\bm{i}}_k \in \gatedom$, an output gate $\check{\bm{o}}_k \in \gatedom$, and a candidate state $\bm{c}_k \in \statedom$:
\begin{subnumcases}{\mathrm{LSTM}\colon\label{eq:lstm_system}}
    \bm{f}_k = \sigma\pr*{W_f \bm{u}_k + R_f \bm{h}_k + \bm{b}_f}, \label{eq:lstm_f} \\
    \check{\bm{i}}_k = \sigma\pr*{W_i \bm{u}_k + R_i \bm{h}_k + \bm{b}_i}, \label{eq:lstm_i} \\
    \check{\bm{o}}_k = \sigma\pr*{W_o \bm{u}_k + R_o \bm{h}_k + \bm{b}_o}, \label{eq:lstm_o} \\
    \bm{c}_k = \phi\pr*{W_c \bm{u}_k + R_c \bm{h}_k + \bm{b}_c}, \label{eq:lstm_c} \\
    \bm{x}_{k+1} = \bm{f}_k \circ \bm{x}_k + \check{\bm{i}}_k \circ \bm{c}_k, \label{eq:lstm_x} \\
    \bm{h}_{k+1} = \check{\bm{o}}_k \circ \phi\pr*{\bm{x}_{k+1}}. \label{eq:lstm_h}
\end{subnumcases}

Two decades later, \citet{choLearningPhraseRepresentations2014} proposed the GRU network, taking a different route.\footnotemark
\footnotetext{Although it may read as a simplified LSTM network, the GRU architecture was designed with little awareness of the LSTM literature~\citep{choNaturalLanguageUnderstanding2015}.}
The GRU architecture keeps a single state, $\bm{h}_k = \bm{x}_k \in \statedom$, updated via a retention (forget) gate $\bm{f}_k \in \gatedom$, a reset gate $\check{\bm{r}}_k \in \gatedom$, and a candidate state $\bm{c}_k \in \statedom$:
\begin{subnumcases}{\mathrm{GRU}\colon\label{eq:gru_system}}
    \bm{f}_k = \sigma\pr*{W_f \bm{u}_k + R_f \bm{h}_k + \bm{b}_f}, \label{eq:gru_f} \\
    \check{\bm{r}}_k = \sigma\pr*{W_r \bm{u}_k + R_r \bm{h}_k + \bm{b}_r}, \label{eq:gru_r} \\
    \bm{c}_k = \phi\pr*{W_c \bm{u}_k + R_c\pr*{\check{\bm{r}}_k \circ \bm{h}_k} + \bm{b}_c}, \label{eq:gru_c} \\
    \bm{x}_{k+1} = \bm{f}_k \circ \bm{x}_k + \pr*{\bm{1} - \bm{f}_k} \circ \bm{c}_k, \label{eq:gru_x} \\
    \bm{h}_{k+1} = \bm{x}_{k+1}. \label{eq:gru_h}
\end{subnumcases}

In both networks, the predicted output $\hat{\bm{y}}_k \in \R^{\ny}$ is obtained from the hidden state through the affine map:
\begin{equation}
\label{eq:affine_output_map}
    \hat{\bm{y}}_k = W_y \bm{h}_k + \bm{b}_y,
\end{equation}
where the output weight matrix $W_y \in \R^{\ny \times n_x}$ and bias vector $\bm{b}_y \in \R^{\ny}$ are collected into $\bparameters_y \coloneq \pc*{W_y, \bm{b}_y}$.

Both networks update the state through the retention term $\bm{f}_k \circ \bm{x}_k$~\citep{greffLSTMSearchSpace2017}, with the forget gate $\bm{f}_k$ computed by the same equation in~\eqref{eq:lstm_f} and~\eqref{eq:gru_f}, and the candidate state $\bm{c}_k$ in~\eqref{eq:lstm_c} and~\eqref{eq:gru_c}.
The two networks differ only in how they combine these terms:
(i)~the GRU network weighs $\bm{c}_k$ by $\bm{1} - \bm{f}_k$, the LSTM network by the independent gate $\check{\bm{i}}_k$;
(ii)~the GRU network scales $\bm{h}_k$ inside the candidate state equation by $\check{\bm{r}}_k$, the LSTM network leaves $\bm{h}_k$ unmodified;
(iii)~the LSTM network maps $\bm{x}_{k+1}$ through $\phi$, the GRU network does not;
and (iv)~the LSTM network scales the result by the output gate $\check{\bm{o}}_k$, the GRU network outputs $\bm{x}_{k+1}$ directly.
We now unify the two networks along these four axes in the coming Section.
%
\section{Generalized Gated RNNs}
\label{s:ggrnn}
%
\begin{figure}[t]
    \vspace*{2pt}
    \centering
    \resizebox{\columnwidth}{!}{%
    \begin{tikzpicture}[
        every node/.style={font=\scriptsize},
        block/.style={rectangle, rounded corners=2pt, draw=gray!85!black, thick, minimum width=0.62cm, minimum height=0.44cm, inner sep=2pt, align=center, fill=gray!5},
        homblock/.style={rectangle, rounded corners=2pt, draw=ggrnn_color, thick, fill=ggrnn_color!8, minimum width=0.92cm, minimum height=0.44cm, inner sep=2pt, align=center},
        eqblock/.style={rectangle, rounded corners=2pt, draw=gray!85!black, thick, fill=gray!4, minimum height=0.44cm, inner sep=2.5pt, align=center},
        had/.style={draw, circle, thick, inner sep=0pt, minimum size=0.26cm, font=\tiny},
        arrow/.style={->, >=stealth, semithick, rounded corners=2pt}
    ]
        \node[block, draw=red!80!black, text=red!80!black, fill=red!6, font=\scriptsize\bfseries] (xk) at (0.00, 2.10) {$\bm{x}_k$};
        \node[block, draw=blue!75!black, text=blue!75!black, fill=blue!6, font=\scriptsize\bfseries] (hk) at (0.00, 1.40) {$\bm{h}_k$};
        \node[block, draw=input_color, text=input_color, fill=input_color!6] (uk) at (0.00, 0.00) {$\bm{u}_k$};

        \node[block] (fk) at (1.60, 2.10) {$\bm{f}_k$};
        \node[homblock] (rk) at (1.60, 1.40) {$\bm{r}_k(\textcolor{ggrnn_color}{\bm{\alpha}_r})$};
        \node[had] (hadr) at (1.60, 0.70) {$\circ$};

        \node[homblock] (ik) at (2.95, 1.40) {$\bm{i}_k(\textcolor{ggrnn_color}{\bm{\alpha}_i})$};
        \node[block] (ck) at (2.95, 0.70) {$\bm{c}_k$};
        \node[homblock] (ok) at (2.95, 0.00) {$\bm{o}_k(\textcolor{ggrnn_color}{\bm{\alpha}_o})$};

        \node[eqblock] (xeq) at (4.75, 2.10) {$\textcolor{red!80!black}{\bm{x}_{k+1}} = \bm{f}_k \circ \textcolor{red!80!black}{\bm{x}_k} + \bm{i}_k \circ \bm{c}_k$};
        \node[homblock] (psi) at (5.20, 1.40) {$\bm{\psi}_{k+1}({\textcolor{ggrnn_color}{\bm{\alpha}_\psi})}$};
        \node[had] (hado) at (6.50, 1.40) {$\circ$};

        \node[block, draw=red!80!black, text=red!80!black, fill=red!6, font=\scriptsize\bfseries] (xkp) at (7.35, 2.10) {$\bm{x}_{k+1}$};
        \node[block, draw=blue!75!black, text=blue!75!black, fill=blue!6, font=\scriptsize\bfseries] (hkp) at (7.35, 1.40) {$\bm{h}_{k+1}$};

        \draw[arrow, draw=red!80!black] (xk.north) -- (0.00, 2.55) -| ([xshift=-0.55cm]xeq.north);

        \draw[arrow, draw=blue!75!black] (hk.east) -- (rk.west);
        \draw[arrow, draw=blue!75!black] (hk.east) -- (0.58, 1.40) |- (fk.west);
        \draw[arrow, draw=blue!75!black] (hk.east) -- (0.58, 1.40) |- (2.25, 1.80) |- ([yshift=+4pt]ik.west);
        \draw[arrow, draw=blue!75!black] (hk.east) -- (0.58, 1.40) |- (hadr.west);
        \draw[arrow, draw=blue!75!black] (hk.east) -- (0.58, 1.40) |- ([yshift=+4pt]ok.west);

        \draw[arrow, draw=input_color] (uk.east) -- (ok.west);
        \draw[arrow, draw=input_color] (uk.east) -- (0.82, 0.00) |- ([yshift=-4pt]fk.west);
        \draw[arrow, draw=input_color] (uk.east) -- (0.82, 0.00) |- ([yshift=-4pt]rk.west);
        \draw[arrow, draw=input_color] (uk.east) -- (2.20, 0.00) |- ([yshift=-4pt]ck.west);
        \draw[arrow, draw=input_color] (uk.east) -- (2.20, 0.00) |- ([yshift=-4pt]ik.west);

        \draw[arrow] (rk.south) -- (hadr.north);
        \draw[arrow] (hadr.east) -- (ck.west);
        \draw[arrow] (fk.east) -- (xeq.west);
        \draw[arrow] (fk.east) -- (2.20, 2.10) |- (ik.west);
        \draw[arrow] (ik.east) -| ([xshift=-0.92cm]xeq.south);
        \draw[arrow] (ck.east) -| ([xshift=-0.65cm]xeq.south);

        \draw[arrow, draw=red!80!black] (xeq.east) -- (xkp.west);
        \draw[arrow, draw=red!80!black] ([xshift=0.45cm]xeq.south) -- (psi.north);
        \draw[arrow] (psi.east) -- (hado.west);
        \draw[arrow] (ok.east) -| (hado.south);
        \draw[arrow, draw=blue!75!black] (hado.east) -- (hkp.west);
    \end{tikzpicture}%
    }
    \vspace{-0.5cm}
    \caption{The GGRNN state update at time $k$.
    Purple blocks highlight the homotopy mechanisms.}
    \label{fig:ggrnn}
\end{figure}
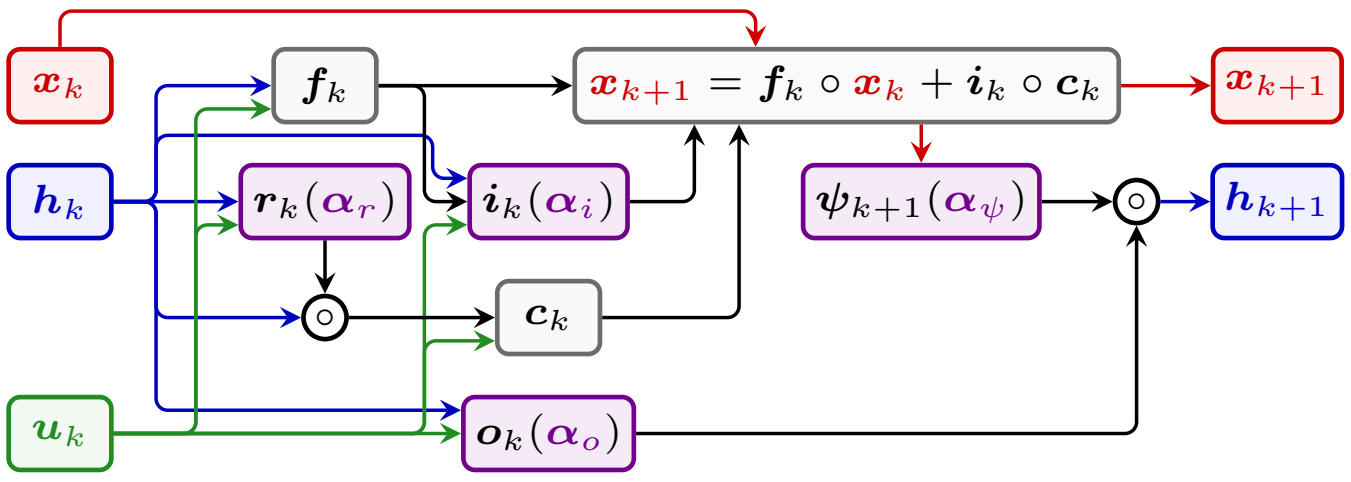

To unify~\eqref{eq:lstm_system} and~\eqref{eq:gru_system} within a single state space model, we introduce four elementwise \emph{homotopies}\footnotemark, one per gating axis, parameterized by the vectors of homotopy parameters $\bm{\alpha}_i, \bm{\alpha}_r, \bm{\alpha}_o, \bm{\alpha}_\psi \in \overline{\gatedom}$, collected into the \emph{homotopy matrix}\arxiv{~(a.k.a.\ the \emph{Steins Gate}\footnotemark)}:
\footnotetext{Given two topological spaces $\mathcal{X}$ and $\mathcal{Y}$, a \emph{homotopy} between two continuous maps $V_0, V_1 : \mathcal{X} \to \mathcal{Y}$ is a family of maps $W(\alpha) : \mathcal{X} \to \mathcal{Y}$, $\alpha \in [0,1]$, such that the associated map $(\alpha, \xi) \mapsto W(\alpha, \xi)$, defined for every $\xi \in \mathcal{X}$ and $\alpha \in [0,1]$, taking values in $\mathcal{Y}$, is continuous with respect to the product topology on $[0,1] \times \mathcal{X}$, with $W(0, \xi) = V_0(\xi)$ and $W(1, \xi) = V_1(\xi)$. If such a family exists, $V_0$ and $V_1$ are called \emph{homotopic} \citep[Ch.~0]{hatcherAlgebraicTopology2002}.}
\arxiv{\footnotetext{\arxiv{In the world-line framing of \emph{Steins;Gate}~(\citealp{steinsgate2009}), every matrix $A$ is a world line interpolating between the GRU and LSTM attractor fields, and the best world lines lie beyond both.}}}
\begin{equation}
\label{eq:homotopy_matrix}
    A \coloneq \ps*{\bm{\alpha}_i,\, \bm{\alpha}_r,\, \bm{\alpha}_o,\, \bm{\alpha}_\psi}^\trans \in \ps{0,1}^{4 \times n_x}.
\end{equation}
The GGRNN computes, at each step $k \in \N$, the gating equations of both networks in~\eqref{eq:lstm_system} and~\eqref{eq:gru_system}, and blends their counterparts elementwise through the homotopies governed by $A$.
Specifically, it evaluates the shared forget gate $\bm{f}_k$ and the three raw gates $(\check{\bm{i}}_k, \check{\bm{r}}_k, \check{\bm{o}}_k)$, blends them into the \emph{effective gates} $\bm{i}_k \in \gatedom$ and $\bm{r}_k, \bm{o}_k \in \left(0,1\right]^{n_x}$, and updates the states according to:
\begin{subnumcases}{\mathrm{GGRNN}\colon\!\label{eq:ggrnn_system}}
    \bm{f}_k = \sigma\pr*{W_f \bm{u}_k + R_f \bm{h}_k + \bm{b}_f}, \label{eq:ggrnn_f} \\
    \check{\bm{i}}_k = \sigma\pr*{W_i \bm{u}_k + R_i \bm{h}_k + \bm{b}_i}, \label{eq:ggrnn_i_raw} \\
    \check{\bm{r}}_k = \sigma\pr*{W_r \bm{u}_k + R_r \bm{h}_k + \bm{b}_r}, \label{eq:ggrnn_r_raw} \\
    \check{\bm{o}}_k = \sigma\pr*{W_o \bm{u}_k + R_o \bm{h}_k + \bm{b}_o}, \label{eq:ggrnn_o_raw} \\
    \bm{i}_k = \pr*{\bm{1} - \bm{\alpha}_i} \circ \pr*{\bm{1} - \bm{f}_k} + \bm{\alpha}_i \circ \check{\bm{i}}_k, \label{eq:ggrnn_i_eff} \\
    \bm{r}_k = \pr*{\bm{1} - \bm{\alpha}_r} \circ \check{\bm{r}}_k + \bm{\alpha}_r \circ \bm{1}, \label{eq:ggrnn_r_eff} \\
    \bm{o}_k = \pr*{\bm{1} - \bm{\alpha}_o} \circ \bm{1} + \bm{\alpha}_o \circ \check{\bm{o}}_k, \label{eq:ggrnn_o_eff} \\
    \bm{c}_k = \phi\pr*{W_c \bm{u}_k + R_c\pr*{\bm{r}_k \circ \bm{h}_k} + \bm{b}_c}, \label{eq:ggrnn_c} \\
    \bm{x}_{k+1} = \bm{f}_k \circ \bm{x}_k + \bm{i}_k \circ \bm{c}_k, \label{eq:ggrnn_x} \\
    \bm{\psi}_{k+1} = \pr*{\bm{1} - \bm{\alpha}_\psi} \circ \bm{x}_{k+1} + \bm{\alpha}_\psi \circ \phi\pr*{\bm{x}_{k+1}}, \label{eq:ggrnn_psi} \\
    \bm{h}_{k+1} = \bm{o}_k \circ \bm{\psi}_{k+1}, \label{eq:ggrnn_h}
\end{subnumcases}
where $\bm{\psi}_{k+1}$ is the \emph{effective state activation}.
\figurename{}~\ref{fig:ggrnn} depicts the resulting architecture.
Since~\eqref{eq:ggrnn_i_eff}--\eqref{eq:ggrnn_o_eff} and~\eqref{eq:ggrnn_psi} combine the GRU and LSTM counterparts of each gating mechanism through convex combinations, they define, for each state element, \emph{linear homotopies} between the corresponding gating maps~\citep[Ch.~0]{hatcherAlgebraicTopology2002}; in particular, the GGRNN dynamics depend continuously on the homotopy matrix $A$.
The predicted output is obtained through~\eqref{eq:affine_output_map}, so that the GGRNN is parameterized by $\bparameters_y$ and $A$ together with $\bParameters_{\mathrm{rnn}} \coloneq \pc*{\bparameters_f, \bparameters_c, \bparameters_i, \bparameters_r, \bparameters_o}$, where $\bparameters_f$ and $\bparameters_c$ are shared by both networks, $\bparameters_r$ contributes only in the GRU network, and $\bparameters_i, \bparameters_o$ only in the LSTM network.

The GGRNN in~\eqref{eq:ggrnn_system} recovers both \emph{endpoint architectures}, i.e., the GRU and LSTM networks, at the homotopy endpoints $A = \bm{0}$ and $A = \bm{1}$.
\begin{proposition}
\label{prop:endpoint_reduction}
    Consider the GGRNN in~\eqref{eq:ggrnn_system}.
    (i)~If $A = \bm{0}$ and $\bm{h}_0 = \bm{x}_0$, then~\eqref{eq:ggrnn_system} reduces to the GRU architecture in~\eqref{eq:gru_system}.
    (ii)~If $A = \bm{1}$, then~\eqref{eq:ggrnn_system} reduces to the LSTM architecture in~\eqref{eq:lstm_system}.
\end{proposition}
\begin{pf}
    (i)~For $A = \bm{0}$,~\eqref{eq:ggrnn_i_eff}--\eqref{eq:ggrnn_o_eff} and~\eqref{eq:ggrnn_psi} yield $\bm{i}_k = \bm{1} - \bm{f}_k$, $\bm{r}_k = \check{\bm{r}}_k$, $\bm{o}_k = \bm{1}$, and $\bm{\psi}_{k+1} = \bm{x}_{k+1}$.
    Together with $\bm{h}_0 = \bm{x}_0$, this gives $\bm{h}_k = \bm{x}_k$ for all $k \in \N$, and the remaining equations on $\bm{x}_k$ reduce to~\eqref{eq:gru_system}.
    (ii)~For $A = \bm{1}$, the same equations yield $\bm{i}_k = \check{\bm{i}}_k$, $\bm{r}_k = \bm{1}$, $\bm{o}_k = \check{\bm{o}}_k$, and $\bm{\psi}_{k+1} = \phi\pr*{\bm{x}_{k+1}}$, coinciding with~\eqref{eq:lstm_system}.
    $\qquad \qquad \qquad \qquad \qquad \qquad \qquad \qquad \qquad \ \ \, \qed$
\end{pf}
%
\section{Training Methodology}
\label{s:training_methodology}
%
\begin{figure}[t]
    \vspace*{2pt}
    \centering
    \resizebox{\columnwidth}{!}{%
    \begin{tikzpicture}[
        every node/.style={font=\scriptsize},
        box/.style={draw, rounded corners=2pt, thick, align=center, inner xsep=3pt, inner ysep=3pt},
        frozen_box/.style={box, draw=cyan!70!blue, fill=cyan!8},
        train_box/.style={box, draw=orange!80!red, fill=orange!8},
        gray_box/.style={box, draw=gray!70, fill=gray!8},
        arrow/.style={->, >=stealth, line width=0.9pt, draw=gray!80!black, rounded corners=2pt}
    ]
        \draw[dashed, draw=orange!70!gray, fill=orange!2, rounded corners=4pt, line width=0.6pt]
            (0.00, 0.00) rectangle (4.10, -3.30);
        \node[font=\scriptsize\bfseries, text=orange!70!black, anchor=north]
            at (2.05,-0.10) {Stage 1: Endpoint Co-Training};

        \node[train_box, text width=3.45cm] (s1_weights) at (2.05,-0.75)  {
            \textcolor{orange!90!red}{\faFire}\ \textbf{Train} $\bParameters$
        };

        \node[gray_box] (s1_gru) at (1.10,-1.50) {
            $\loss{GRU}\pr*{\bParameters}$
        };
        \node[gray_box] (s1_lstm) at (3.00,-1.50) {
            $\loss{LSTM}\pr*{\bParameters}$
        };

        \node[gray_box, text width=3.7cm] (s1_loss) at (2.05,-2.25) {
            $\min_{\bParameters} \loss{S1}\pr*{\bParameters}$
        };

        \node[gray_box] (s1_result) at (2.05,-3.00) {
            $\bParameters^\star$
        };

        \draw[arrow] (s1_gru.north |- s1_weights.south) -- (s1_gru.north);
        \draw[arrow] (s1_lstm.north |- s1_weights.south) -- (s1_lstm.north);
        \draw[arrow] (s1_gru.south) -- (s1_gru.south |- s1_loss.north);
        \draw[arrow] (s1_lstm.south) -- (s1_lstm.south |- s1_loss.north);
        \draw[arrow] (s1_loss.south) -- (s1_result.north);

        \draw[dashed, draw=ggrnn_color!80!gray, fill=ggrnn_color!3, rounded corners=4pt, line width=0.6pt]
            (4.30,0.00) rectangle (8.40,-3.30);
        \node[font=\scriptsize\bfseries, text=ggrnn_color, anchor=north]
            at (6.35,-0.10) {Stage 2: Homotopy Optimization};

        \node[frozen_box, text width=3.45cm] (s2_frozen) at (6.35,-0.75) {
            \textcolor{cyan!70!blue}{\faSnowflake}\ \textbf{Freeze} $\bParameters^\star$
        };

        \node[train_box, text width=3.45cm] (s2_alpha) at (6.35,-1.50) {
            \textcolor{orange!90!red}{\faFire}\ \textbf{Train} $H: A(H)=\sigma(H)$
        };

        \node[gray_box, text width=3.45cm] (s2_loss) at (6.35,-2.25) {
            $\min_{H} \loss{S2}\pr{H}$
        };

        \node[gray_box] (s2_result) at (6.35,-3.00) {
            $H^\star$
        };

        \draw[arrow] (s2_frozen.south) -- (s2_alpha.north);
        \draw[arrow] (s2_alpha.south) -- (s2_loss.north);
        \draw[arrow] (s2_loss.south) -- (s2_result.north);

        \draw[arrow]
            (s1_result.east)
            -- (4.20,-3.00)
            -- (4.20,-0.75)
            -- (s2_frozen.west);
    \end{tikzpicture}%
    }
    \vspace{-0.5cm}
    \caption{Training methodology: Stage~1 co-trains $\bParameters$ at the GRU ($A=\bm{0}$) and LSTM ($A=\bm{1}$) endpoints; Stage~2 freezes $\bParameters^\star$ and optimizes only the homotopy logits $H$.
    \textcolor{orange!90!red}{\smash{\faFire}} and \textcolor{cyan!70!blue}{\smash{\faSnowflake}} denote trainable and frozen parameters; stars indicate validation-selected parameters.
    }
    \label{fig:training}
\end{figure}
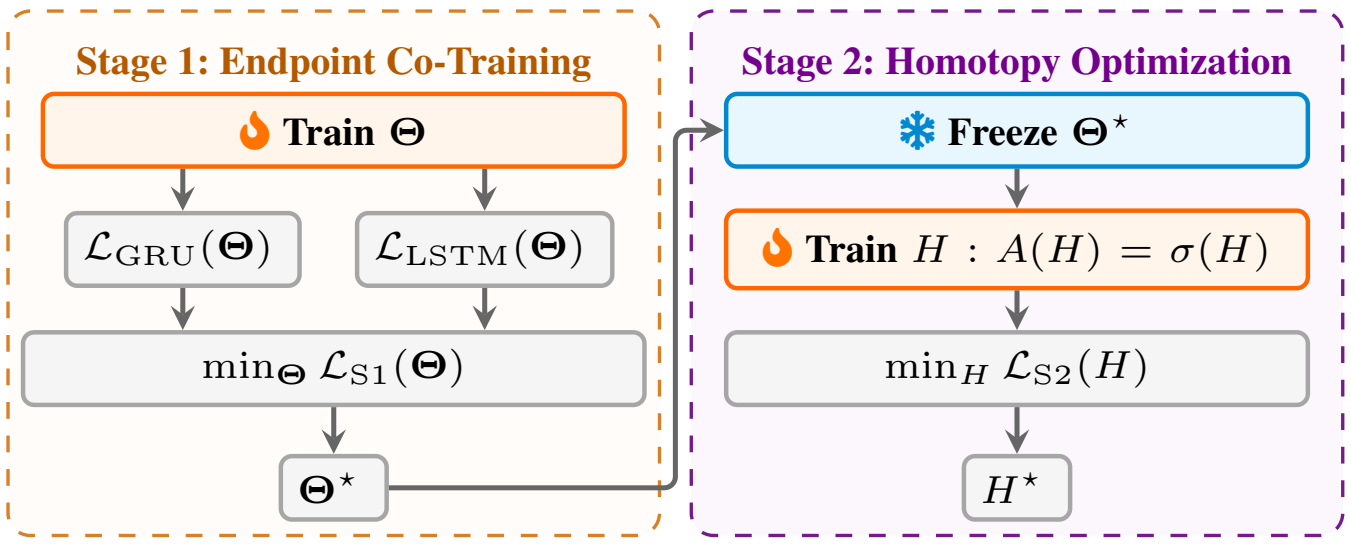

In principle, we could directly identify the GGRNN in~\eqref{eq:ggrnn_system} end-to-end by optimizing the parameters $\pc*{\bParameters_{\mathrm{rnn}}, \bparameters_y}$ and the homotopy matrix $A$ jointly from random initialization.
This approach, however, is poorly conditioned: by~\eqref{eq:ggrnn_i_eff}--\eqref{eq:ggrnn_o_eff} and~\eqref{eq:ggrnn_psi}, the gradients with respect to $A$ are weighed by the differences between the GRU and LSTM counterparts of each gating mechanism, e.g., $\check{\bm{i}}_k - \pr{\bm{1} - \bm{f}_k}$ in~\eqref{eq:ggrnn_i_eff}; as long as the parameters are still random, such differences are uninformative about which endpoint fits the data better~\citep[Section~8.2]{goodfellowDeepLearning2016}.
We address this difficulty with the two-stage methodology illustrated in \figurename{}~\ref{fig:training}: in Stage~1 (Section~\ref{ss:stage1}), we co-train the network parameters at the two endpoint architectures; in Stage~2 (Section~\ref{ss:stage2}), we freeze those parameters and optimize the homotopy matrix $A$ alone, reparameterized through unconstrained homotopy logits $H \in \R^{4 \times \nx}$ with $A\pr{H} = \sigma\pr{H}$.
%
\subsection{State Initialization}
\label{ss:state_init}
%
Training over the full dataset $\ds$ in Section~\ref{ss:problem_statement} requires unrolling and storing several time steps, which is largely inefficient; we therefore train via truncated backpropagation through time on windows of $\ds$.
Specifically, we partition $\ds$ into $N_w \in \Zpos$ training windows of $P \in \Zpos$ prediction steps: the window with start index $k_0 \in \mathcal{K}$ collects the samples $\pc*{\pr{\bm{u}_k, \bm{y}_k}}_{k=k_0}^{k_0+P-1}$, where $\mathcal{K} \subset \Zpos$ collects the $N_w$ start indices.
We estimate the initial state $\ps{\bm{x}_{k_0}^\top,\bm{h}_{k_0}^\top}^\top$ at the start of each window with a subspace encoder~\citep{beintemaDeepSubspaceEncoders2023} with parameters $\bparameters_e$, which maps the past input--output samples $\pc*{\pr{\bm{u}_k, \bm{y}_k}}_{k=k_0-T_{\mathrm{past}}}^{k_0-1}$ of length $T_{\mathrm{past}} \in \Zpos$ to raw preactivations.
We depart from the original formulation in two adjustments needed for the GGRNN: the encoder produces $2n_x$ raw preactivations $\ps{\bm{z}_{\mathrm{x},k_0}^\trans, \bm{z}_{o,k_0}^\trans}^\trans \in \R^{2n_x}$ rather than an $n_x$-dimensional state estimate, and the window initial states 
$\bm{x}_{k_0}$ and $\bm{h}_{k_0}$ are reconstructed by setting $\check{\bm{o}}_{k_0} = \sigma\pr*{\bm{z}_{o,k_0}}$ and:
\begin{equation}
    \label{eq:subnet_init}
    \bm{x}_{k_0} = \phi\pr*{\bm{z}_{\mathrm{x},k_0}}, \qquad \bm{h}_{k_0} = \bm{o}_{k_0} \circ \bm{\psi}_{k_0},
\end{equation}
where $\bm{o}_{k_0}$ and $\bm{\psi}_{k_0}$ are induced by $\check{\bm{o}}_{k_0}$ and $\bm{x}_{k_0}$ via~\eqref{eq:ggrnn_o_eff} and~\eqref{eq:ggrnn_psi}, respectively, the hyperbolic tangent guarantees $\bm{x}_{k_0} \in \statedom$, and $\bm{o}_{k_0} \in \left(0,1\right]^{n_x}$ with $\bm{\psi}_{k_0} \in \statedom$ guarantee $\bm{h}_{k_0} \in \statedom$ as well.
At the GRU endpoint $A = \bm{0}$, the output gate is bypassed, i.e. $\bm{o}_{k_0} = \bm{1}$, only $\bm{z}_{\mathrm{x},k_0}$ is used, and~\eqref{eq:subnet_init} yields $\bm{h}_{k_0} = \bm{x}_{k_0}$, matching the endpoint condition (i) of Proposition~\ref{prop:endpoint_reduction}; at the LSTM endpoint $A = \bm{1}$, both preactivations are used.
%
\subsection{Stage 1: Endpoint Co-Training}
\label{ss:stage1}
%
Let $\bParameters \coloneq \pc*{\bparameters_e, \bParameters_{\mathrm{rnn}}, \bparameters_y}$ collect the parameters of the encoder, the GGRNN, and the output map, and let
\begin{equation}
    \label{eq:mse_loss_general}
    \mathcal{L}\pr*{\bParameters, A} \coloneq \frac{1}{N_w P \ny} \sum_{k_0 \in \mathcal{K}} \sum_{p=0}^{P-1} \norm*{\bm{y}_{k_0+p}-\hat{\bm{y}}_{k_0+p}\pr*{\bParameters,A}}^2
\end{equation}
denote the simulation Mean Squared Error (MSE) over the training windows of Section~\ref{ss:state_init}, where, within each window, the prediction $\hat{\bm{y}}_{k}\pr*{\bParameters,A}$ is obtained by simulating the GGRNN in~\eqref{eq:ggrnn_system} with homotopy matrix $A$ from the initial states of Section~\ref{ss:state_init}, with the outputs given by~\eqref{eq:affine_output_map}.
In Stage~1, we evaluate~\eqref{eq:mse_loss_general} in parallel at the two endpoints, yielding the architecture losses $\loss{GRU}\pr*{\bParameters} \coloneq \mathcal{L}\pr*{\bParameters, \bm{0}}$ and $\loss{LSTM}\pr*{\bParameters} \coloneq \mathcal{L}\pr*{\bParameters, \bm{1}}$.
With $\lambda \in \R_{\geq 0}$, we solve:
\begin{equation}
    \label{eq:stage1_objective}
    \min_{\bParameters} \; \loss{S1}\pr*{\bParameters} \coloneq \loss{GRU}\pr*{\bParameters} + \loss{LSTM}\pr*{\bParameters} + \lambda R\pr*{\bParameters},
\end{equation}
where $R\pr*{\bParameters}$ is the sum of the squared Frobenius norms of the weight matrices of $\bParameters$, excluding the biases.
We additionally hold out a validation dataset, on which $\loss{GRU}^{\mathrm{val}}\pr*{\bParameters}$ and $\loss{LSTM}^{\mathrm{val}}\pr*{\bParameters}$ denote the simulation MSE of~\eqref{eq:mse_loss_general}, evaluated at $A=\bm{0}$ and $A=\bm{1}$, respectively; the evaluation protocol is specified in Section~\ref{ss:training_setup}.
We retain the parameters $\bParameters^\star$ that minimize the average endpoint validation MSE, $\frac{1}{2}\pr*{\loss{GRU}^{\mathrm{val}}\pr*{\bParameters}+\loss{LSTM}^{\mathrm{val}}\pr*{\bParameters}}$, and hold them fixed in Stage~2.
\begin{remark}
    In the data-fitting terms of~\eqref{eq:stage1_objective}, the blocks shared across the two endpoints $\pr*{\bparameters_e, \bparameters_f, \bparameters_c, \bparameters_y}$ receive summed gradients from both endpoint simulations, while $\bparameters_r$ affects exclusively $\loss{GRU}\pr*{\bParameters}$ and $\pr{\bparameters_i, \bparameters_o}$ affects exclusively $\loss{LSTM}\pr*{\bParameters}$.
    This sharing acts as a regularizer~\citep[Section~7.9]{goodfellowDeepLearning2016}: the shared parameters must remain effective for both endpoint architectures rather than specialize to either one, and the latent representation stays compatible with the two distinct state updates.
    %
\end{remark}
%
\subsection{Stage 2: Homotopy Optimization}
\label{ss:stage2}
%
In Stage~2, all network parameters $\bParameters^\star$ are frozen, and we optimize only the homotopy logits $H$.
Specifically, we solve:
\begin{equation*}
    \min_{H} \; \loss{S2}\pr{H} \coloneq \mathcal{L}\pr*{\bParameters^\star, A\pr{H}},
\end{equation*}
and retain the logits $H^\star$ with the lowest validation MSE on the same validation dataset, setting $A^\star = A\pr{H^\star}$.
\begin{remark}
    The two-stage training procedure returns three configurations: the two endpoint LSTM and GRU models at the frozen Stage~1 with parameters $\bParameters^\star$, and the Stage~2 optimum GGRNN with homotopy matrix $A^\star$ and $\bParameters^\star$ fixed.
    No separate training runs are required for the two endpoint models.
\end{remark}
%
\section{Results on Nonlinear Benchmarks}
\label{s:results}
%
\begin{table}[!htbp]
    \centering
    \caption{Dimensions and training hyperparameters.}
    \vspace{-0.2cm}
    \label{tab:training_settings}
    \settablestretch
    \settablefontsize
    \setlength{\tabcolsep}{3pt}
    \begin{tabular*}{\columnwidth}
        {@{\extracolsep{\fill}} l c c @{}}
        \toprule
        \textbf{Setting} & \textbf{W--HB} & \textbf{N--DB} \\
        \midrule
        Dimensions $(n_u,n_y,n_x)$ & $(1,1,6)$ & $(4,12,64)$ \\
        Encoder $(T_{\mathrm{past}}, \nenc)$ & $(80, 32)$ & $(20, 64)$ \\
        Windows ($P$, stride) & $(250, 8)$ & $(100, 4)$ \\
        Learning rate (initial, final) & $\pr{3.5\times10^{-3},\, 5\times10^{-5}}$ & $\pr{10^{-3},\, 2\times10^{-5}}$ \\
        Epochs (Stages 1, 2) & $(3000,1000)$ & $(2000,2000)$ \\
        Regularization penalty $\lambda$ & $10^{-6}$ & $10^{-6}$ \\
        \bottomrule
    \end{tabular*}
\end{table}

We evaluate the proposed methodology on two public nonlinear system identification benchmarks from the collection of~\citet{schoukensBenchmarkingNonlinearSystem2026}: the Wiener--Hammerstein and the Nano--Drone benchmarks.
%
\subsection{Wiener--Hammerstein Benchmark}
\label{ss:wh_benchmark}
%
The Wiener--Hammerstein Benchmark (W--HB)~\citep{schoukensWienerHammersteinBenchmark2009} is a single-input single-output electronic circuit, consisting of a static saturation nonlinearity between two third-order linear filters.
The scalar input $u_k$ and output $y_k$ are voltages measured in $\mathrm{V}$.
The circuit is excited by filtered Gaussian noise sampled at $51.2\,\mathrm{kHz}$, and simulation accuracy is reported through the output Root MSE (RMSE) in $\mathrm{mV}$.
We follow the established benchmark separation between the $100\,000$-sample estimation record and the independent test record, reserving the first $80\,000$ estimation samples for training and the remaining $20\,000$ for validation.
Test performance is evaluated in simulation, with the initial state estimated from the past input--output window as in Section~\ref{ss:state_init}.

We compare against the independently trained GRU and LSTM baselines of~\citet{champneysBaselineResultsSelected2024}, whose implementations follow the architectures reviewed in Section~\ref{ss:gated_rnns} with a state dimension of $\nx = 32$.
The baselines use the lagged-input vector $\ps*{u_k,\dots,u_{k-31}}^\trans \in \R^{32}$ as network input, whereas our models simply take the last input $u_k$ alone.
%
\subsection{Nano--Drone Benchmark}
\label{ss:nanodrone_benchmark}
%
The Nano--Drone Benchmark (N--DB)~\citep{busettoNonlinearSystemIdentification2026} is a Crazyflie nano-quadrotor, i.e., a MIMO system with coupled translational and rotational dynamics with signals sampled at $100\,\mathrm{Hz}$.
The input $\bm{u}_k\in\R^4$ contains the four motor signals.
Following~\citet{busettoNonlinearSystemIdentification2026}, we use $12$ outputs: world-frame position $\bm{p}_k=[p_{\mathrm{x},k},p_{\mathrm{y},k},p_{\mathrm{z},k}]^\trans$ [$\mathrm{m}$], world-frame linear velocity $\bm{v}_k=[v_{\mathrm{x},k},v_{\mathrm{y},k},v_{\mathrm{z},k}]^\trans$ [$\mathrm{m/s}$], attitude expressed as a rotation vector $\bm{r}_k=[r_{\mathrm{x},k},r_{\mathrm{y},k},r_{\mathrm{z},k}]^\trans$ [$\mathrm{rad}$], and body-frame angular velocity $\bm{\omega}_k=[\omega_{\mathrm{x},k},\omega_{\mathrm{y},k},\omega_{\mathrm{z},k}]^\trans$ [$\mathrm{rad/s}$].
Following the benchmark protocol, Square, Random, and Chirp input signals provide the identification trajectories, while the three Melon flights are reserved for testing.
For each identification trajectory, we use three runs for training and the fourth for validation, yielding nine training and three validation flights.
Test performance is evaluated over rolling open-loop simulation windows of $T = 50$ steps ($0.5\,\mathrm{s}$): for each window origin $k$, the initial state is estimated from the past input--output window as in Section~\ref{ss:state_init}, and the model simulates the $T$ steps ahead.

Following~\citet{busettoNonlinearSystemIdentification2026}, test performance is reported for four output groups, indexed by $\examplesymbol \in \pc{p, v, r, \omega}$: position, linear velocity, attitude, and angular velocity, respectively.
As a shorthand, we write $\bm{y}_k^{\examplesymbol}$ for the measured output block and $\hat{\bm{y}}_k^{\examplesymbol}\pr*{\bParameters^{\star}, A^{\star}}$ for the corresponding block of the predicted output.
For each test flight $j \in \iset{1}{n_{\mathrm{f}}}$, with $n_{\mathrm{f}} = 3$, let $\mathcal{W}_j \subset \N$ collect the origins $k$ of its rolling simulation windows, so that each $k \in \mathcal{W}_j$ and $t \in \iset{1}{T}$ identify the prediction $\hat{\bm{y}}^\examplesymbol_{k+t}\pr*{\bParameters^{\star}, A^{\star}}$.
For each prediction step $t \in \iset{1}{T}$, we average the resulting errors over all window origins within each test flight, and then across the test flights, yielding the Mean Absolute Error (MAE):
\begin{equation}
    \label{eq:nanodrone_mae}
    \mathrm{MAE}_{\examplesymbol}\pr{t} \! \coloneq \! \frac{1}{n_{\mathrm{f}}} \sum_{j=1}^{n_{\mathrm{f}}} \frac{1}{\abs{\mathcal{W}_j}} \sum_{k \in \mathcal{W}_j} \norm*{\hat{\bm{y}}_{k+t}^{\examplesymbol}\pr*{\bParameters^{\star}, A^{\star}} \! - \! \bm{y}_{k+t}^{\examplesymbol}}.
\end{equation}
The attitude error is defined analogously, with the relative rotation angle [$\mathrm{rad}$] between the rotations induced by $\hat{\bm{y}}_{k+t}^{r}$ and $\bm{y}_{k+t}^{r}$ in place of the norm.
Besides the values of~\eqref{eq:nanodrone_mae} at $t = 1, 10, 50$, we report the cumulative simulation error:
\begin{equation*}
    E_{\examplesymbol} \coloneq \sum_{t=1}^{T} \mathrm{MAE}_{\examplesymbol}\pr{t}.
\end{equation*}
We compare against the reference models of~\citet{busettoNonlinearSystemIdentification2026}: the first-principles Physics model, the grey-box Hybrid model, and the black-box residual models Res-MLP and Res-LSTM.
The two neural network baselines predict output increments that are accumulated recursively over the simulation horizon.
Res-LSTM uses a state dimension of $\nx=64$, providing an independently trained LSTM reference.
\begin{table}[!htbp]
    \centering
    \caption{Simulation results on the W--HB against the baselines in~\citet{champneysBaselineResultsSelected2024}.
    Bold denotes the lowest RMSE.}
    \vspace{-0.2cm}
    \label{tab:results_wh}
    \settablestretch
    \settablefontsize
    \setlength{\tabcolsep}{6pt}
    \begin{tabular*}{\columnwidth}{@{\extracolsep{\fill}} l l c @{}}
        \toprule
        \textbf{Source} & \textbf{Model} & \textbf{RMSE [mV]} \\
        \midrule
        \multirow{2}{*}{\makecell[l]{Baselines ($n_x=32$) \\ \citep{champneysBaselineResultsSelected2024}}}
        & Baseline GRU                          & $3.97$ \\
        & Baseline LSTM                         & $4.45$ \\
        \midrule
        \multirow{3}{*}{\makecell[l]{Proposed ($n_x=6$)}}
        & GRU ($A = \bm{0}$)                    & $3.94$ \\
        & LSTM ($A = \bm{1}$)                   & $3.71$ \\
        & GGRNN ($A^\star$)                     & $\mathbf{3.50}$ \\
        \bottomrule
    \end{tabular*}
    \vspace{1.5pt}
    \raggedright
\end{table}
\begin{table*}[!htbp]
    \centering
    \caption{N--DB test MAEs at selected prediction steps and cumulative errors $E_{\examplesymbol}$ over $50$ steps.
    Reference results are from~\citet{busettoNonlinearSystemIdentification2026}.
    Grayed-out columns report per-step MAEs; bold denotes the lowest cumulative error in each group.}
    \vspace{-6pt}
    \label{tab:results_nanodrone}
    \settablestretch
    \settablefontsize
    \setlength{\tabcolsep}{2.2pt}
    \begin{tabular*}{\textwidth}{@{\extracolsep{\fill}} l 
        >{\color{black!50}}c >{\color{black!50}}c >{\color{black!50}}c c 
        >{\color{black!50}}c >{\color{black!50}}c >{\color{black!50}}c c 
        >{\color{black!50}}c >{\color{black!50}}c >{\color{black!50}}c c 
        >{\color{black!50}}c >{\color{black!50}}c >{\color{black!50}}c c @{}}
        \toprule
        & \multicolumn{4}{c}{Position MAE$_p$ [m]}
        & \multicolumn{4}{c}{Velocity MAE$_v$ [m/s]}
        & \multicolumn{4}{c}{Attitude MAE$_r$ [rad]}
        & \multicolumn{4}{c}{Angular velocity MAE$_\omega$ [rad/s]} \\
        \cmidrule(lr){2-5} \cmidrule(lr){6-9} \cmidrule(lr){10-13} \cmidrule(lr){14-17}
        \textbf{Model} 
        & $t=1$ & $t=10$ & $t=50$ & $E_p$
        & $t=1$ & $t=10$ & $t=50$ & $E_v$
        & $t=1$ & $t=10$ & $t=50$ & $E_r$
        & $t=1$ & $t=10$ & $t=50$ & $E_\omega$\\
        \midrule
        Physics     & 0.0013 & 0.0126 & 0.1269 & \textbf{2.32}   & 0.0080 & 0.0570 & 0.5781 & 10.62           & 0.0011 & 0.0205 & 0.2544 & 5.10            & 0.0796 & 0.3596 & 0.8837 & 29.09 \\
        Hybrid      & 0.0016 & 0.0166 & 0.1119 & 2.36            & 0.0092 & 0.0613 & 0.5556 & \textbf{10.40}  & 0.0027 & 0.0376 & 0.2306 & 6.15            & 0.0912 & 0.4880 & 0.5979 & \textbf{28.99} \\
        \midrule
        Res-MLP     & 0.0032 & 0.0305 & 0.1712 & 4.05            & 0.0116 & 0.0890 & 0.5720 & 12.58           & 0.0022 & 0.0331 & 0.2268 & 6.06            & 0.1138 & 0.5949 & 0.9005 & 35.57 \\
        Res-LSTM    & 0.0079 & 0.0390 & 0.2572 & 5.97            & 0.0247 & 0.1175 & 0.7407 & 16.72           & 0.0066 & 0.0372 & 0.2325 & 5.65            & 0.1021 & 0.4292 & 1.2407 & 35.84 \\
        \midrule
        GRU ($A=\bm{0}$)      & 0.0086 & 0.0440 & 0.1815 & 4.75 & 0.0226 & 0.1133 & 0.4996 & 12.40 & 0.0053 & 0.0412 & 0.1700 & 4.90          & 0.1015 & 0.4688 & 0.8844 & 32.23 \\
        LSTM ($A=\bm{1}$)     & 0.0077 & 0.0436 & 0.1951 & 5.02 & 0.0227 & 0.1099 & 0.4981 & 12.07 & 0.0055 & 0.0391 & 0.1743 & 4.84          & 0.0952 & 0.4213 & 0.8654 & 29.64 \\
        GGRNN ($A^\star$)     & 0.0078 & 0.0427 & 0.1883 & 4.85 & 0.0224 & 0.1073 & 0.4832 & 11.75 & 0.0053 & 0.0386 & 0.1647 & \textbf{4.65} & 0.0951 & 0.4310 & 0.8181 & 29.34 \\
        \bottomrule
    \end{tabular*}
    \raggedright
\end{table*}
%
\subsection{Training Setup}
\label{ss:training_setup}
%
For GGRNN training, inputs and outputs are standardized using only the training set statistics.
All recurrent models use a single recurrent layer.
The state encoder is a perceptron with a single hidden layer of $\nenc$ rectified linear units and a linear output layer producing the $2\nx$ preactivations of Section~\ref{ss:state_init}.

For the W--HB, predictions use the direct output map in~\eqref{eq:affine_output_map}.
For the N--DB, following the residual prediction strategy of~\citet{busettoNonlinearSystemIdentification2026}, we use the residual output map $\hat{\bm{y}}_{k+1}=\hat{\bm{y}}_k+W_y\bm{h}_{k+1}+\bm{b}_y$, initialized with the last measured output in the encoder history.

Stages~1 and~2 use Adam with global gradient-norm clipping at $1.5$ and a cosine learning-rate schedule decaying between its initial and final values, initializing a fresh optimizer at each stage.
Training windows of length $P$ are shuffled into mini-batches of $256$ windows; Stage~1 evaluates both LSTM and GRU networks on the same mini-batch.
Stage~2 optimizes the unconstrained homotopy logits $H$, initialized as $H_0 = \sigma^{-1}\pr{A_0}$ entrywise, where $\sigma^{-1}$ is the inverse sigmoid and the entries of $A_0$ are sampled independently from a uniform distribution on $\ps{0.05,\,0.95}$.
Validation is performed every $50$ epochs and at the final epoch.
Stage~2 stops early if the validation MSE does not improve for $200$ consecutive epochs.
The benchmark-specific model dimensions and training hyperparameters are summarized in Table~\ref{tab:training_settings}; implementation details are available in the replication package~\citep{decarliGGRNNs2026}.

We select the parameters $\bParameters^\star$ and $H^\star$ of Sections~\ref{ss:stage1} and~\ref{ss:stage2} by their validation losses, computed on standardized outputs: in simulation for the W--HB and over $P$-step validation windows for N--DB.
For each benchmark, we perform $10$ two-stage runs differing only in the random seed of the parameter initializations, and select the run with the lowest Stage~2 validation MSE for test evaluation.
The reported GRU and LSTM endpoints use the shared Stage~1 parameters $\bParameters^{\star}$ of that same selected run.

Across ten runs, training took approximately $46$~min per W--HB run and $21$~min per N--DB run.
Stage~1 accounted for about $40$ and $18$~min, respectively, while Stage~2 took about $6$ and $3$~min, respectively.\footnotemark
\footnotetext{Experiments use JAX/Equinox on a workstation equipped with an AMD Ryzen 9 7900X CPU, $32$~GB RAM, and an NVIDIA GeForce RTX 4070 Ti SUPER GPU with $16$~GB memory.}
%
\subsection{Numerical Results}
\label{ss:numerical_results}
%
\begin{figure*}[!htbp]
    \vspace*{2pt}
    \centering
    \begin{tikzpicture}
        \begin{groupplot}[
            group style={
                group size=4 by 1,
                horizontal sep=0.95cm,
                vertical sep=0.22cm,
                xlabels at=edge bottom,
                xticklabels at=edge bottom,
            },
            width=0.19\textwidth,
            height=1.5cm,
            scale only axis,
            xmin=1, xmax=50,
            xlabel={Prediction steps $t$},
            xtick={1, 10, 20, 30, 40, 50},
            grid=major,
            grid style={dashed, gray!30},
            tick label style={font=\scriptsize},
            label style={font=\scriptsize},
            ylabel style={yshift=-4pt},
            xlabel style={yshift=2pt},
        ]
        \nextgroupplot[
            ylabel={$\mathrm{MAE}_{p}(t)$ [m]},
            ymin=0, ymax=0.28,
            ytick={0, 0.1, 0.2},
            legend columns=7,
            legend style={
                at={(2.5,1.15)},
                anchor=south,
                font=\tiny,
                draw=black!40,
                fill=white,
                inner sep=1.2pt,
                column sep=2.5pt,
            },
        ]
            \addplot[black, dotted, line width=0.7pt] table[x=t, y=physP] {data/nanodrone_horizon.dat};
\addlegendentry{Physics}
\addplot[blue!70!black, densely dashed, line width=0.7pt] table[x=t, y=hybridP] {data/nanodrone_horizon.dat};
\addlegendentry{Hybrid}
\addplot[mlp_color, densely dashed, line width=0.7pt] table[x=t, y=mlpP] {data/nanodrone_horizon.dat};
\addlegendentry{Res-MLP}
\addplot[lstm_color, densely dotted, line width=0.7pt] table[x=t, y=reslstmP] {data/nanodrone_horizon.dat};
\addlegendentry{Res-LSTM}
\addplot[color=gru_color, dashed, line width=0.8pt] table[x=t, y=gruP] {data/nanodrone_horizon.dat};
            \addlegendentry{GRU ($A=\bm{0}$)}
            \addplot[color=lstm_color, dashdotted, line width=0.8pt] table[x=t, y=lstmP] {data/nanodrone_horizon.dat};
            \addlegendentry{LSTM ($A=\bm{1}$)}
            \addplot[color=ggrnn_color, solid, line width=0.85pt] table[x=t, y=ggrnnP] {data/nanodrone_horizon.dat};
            \addlegendentry{GGRNN ($A^\star$)}
        \nextgroupplot[
            ylabel={$\mathrm{MAE}_{v}(t)$ [m/s]},
            ymin=0, ymax=0.80,
            ytick={0, 0.3, 0.6},
        ]
            \addplot[black, dotted, line width=0.7pt] table[x=t, y=physV] {data/nanodrone_horizon.dat};
\addplot[blue!70!black, densely dashed, line width=0.7pt] table[x=t, y=hybridV] {data/nanodrone_horizon.dat};
\addplot[mlp_color, densely dashed, line width=0.7pt] table[x=t, y=mlpV] {data/nanodrone_horizon.dat};
\addplot[lstm_color, densely dotted, line width=0.7pt] table[x=t, y=reslstmV] {data/nanodrone_horizon.dat};
\addplot[color=gru_color, dashed, line width=0.8pt] table[x=t, y=gruV] {data/nanodrone_horizon.dat};
            \addplot[color=lstm_color, dashdotted, line width=0.8pt] table[x=t, y=lstmV] {data/nanodrone_horizon.dat};
            \addplot[color=ggrnn_color, solid, line width=0.85pt] table[x=t, y=ggrnnV] {data/nanodrone_horizon.dat};
        \nextgroupplot[
            ylabel={$\mathrm{MAE}_{r}(t)$ [rad]},
            ymin=0, ymax=0.28,
            ytick={0, 0.1, 0.2},
        ]
            \addplot[black, dotted, line width=0.7pt] table[x=t, y=physR] {data/nanodrone_horizon.dat};
\addplot[blue!70!black, densely dashed, line width=0.7pt] table[x=t, y=hybridR] {data/nanodrone_horizon.dat};
\addplot[mlp_color, densely dashed, line width=0.7pt] table[x=t, y=mlpR] {data/nanodrone_horizon.dat};
\addplot[lstm_color, densely dotted, line width=0.7pt] table[x=t, y=reslstmR] {data/nanodrone_horizon.dat};
\addplot[color=gru_color, dashed, line width=0.8pt] table[x=t, y=gruR] {data/nanodrone_horizon.dat};
            \addplot[color=lstm_color, dashdotted, line width=0.8pt] table[x=t, y=lstmR] {data/nanodrone_horizon.dat};
            \addplot[color=ggrnn_color, solid, line width=0.85pt] table[x=t, y=ggrnnR] {data/nanodrone_horizon.dat};
        \nextgroupplot[
            ylabel={$\mathrm{MAE}_{\omega}(t)$ [rad/s]},
            ymin=0, ymax=1.35,
            ytick={0, 0.5, 1.0},
        ]
            \addplot[black, dotted, line width=0.7pt] table[x=t, y=physW] {data/nanodrone_horizon.dat};
\addplot[blue!70!black, densely dashed, line width=0.7pt] table[x=t, y=hybridW] {data/nanodrone_horizon.dat};
\addplot[mlp_color, densely dashed, line width=0.7pt] table[x=t, y=mlpW] {data/nanodrone_horizon.dat};
\addplot[lstm_color, densely dotted, line width=0.7pt] table[x=t, y=reslstmW] {data/nanodrone_horizon.dat};
\addplot[color=gru_color, dashed, line width=0.8pt] table[x=t, y=gruW] {data/nanodrone_horizon.dat};
            \addplot[color=lstm_color, dashdotted, line width=0.8pt] table[x=t, y=lstmW] {data/nanodrone_horizon.dat};
            \addplot[color=ggrnn_color, solid, line width=0.85pt] table[x=t, y=ggrnnW] {data/nanodrone_horizon.dat};
        \end{groupplot}
    \end{tikzpicture}
    \vspace{-0.4cm}
    \caption{N--DB test MAE versus prediction steps $t$ for position, linear velocity, attitude, and angular velocity.
    Reference curves are reconstructed from the predictions released by~\citet{busettoNonlinearSystemIdentification2026}.}
    \vspace{-6pt}
    \label{fig:nanodrone_horizon}
\end{figure*}
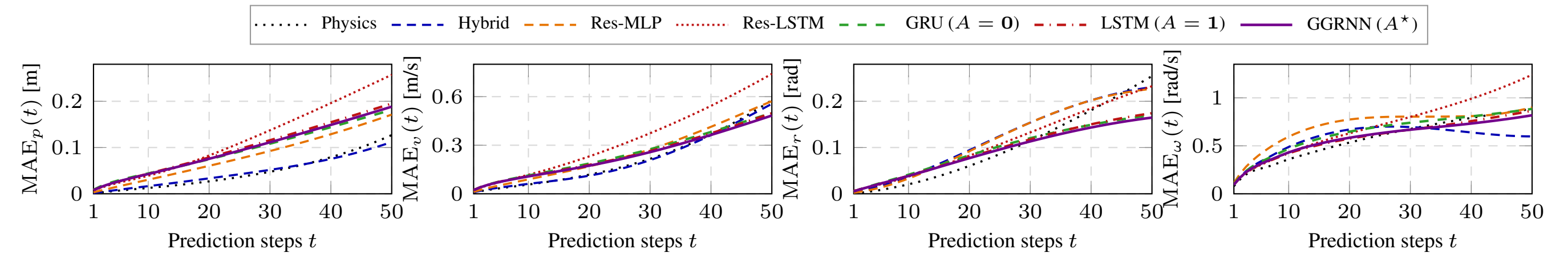

\providecolor{mlp_color}{RGB}{230, 115, 0}
\begin{figure*}[!htbp]
    \vspace*{2pt}
    \centering
    \begin{tikzpicture}
        \begin{groupplot}[
            group style={
                group size=3 by 4,
                horizontal sep=1.15cm,
                vertical sep=0.20cm,
                xlabels at=edge bottom,
                xticklabels at=edge bottom,
            },
            width=4.60cm,
            height=1.25cm,
            scale only axis,
            xmin=20.5, xmax=25.5,
            xtick={21, 22, 23, 24, 25},
            grid=both,
            grid style={line width=0.2pt, draw=gray!20},
            label style={font=\scriptsize},
            tick label style={font=\tiny},
            ylabel style={font=\scriptsize, at={(axis description cs:-0.12, 0.5)}, anchor=south},
            xlabel style={yshift=2pt},
        ]

            \nextgroupplot[
                ylabel={$p_\mathrm{x}$ [m]},
                ymin=-0.8, ymax=0.8,
                ytick={-0.5, 0, 0.5},
                legend columns=4,
                legend style={
                    at={(1.75, 1.25)},
                    anchor=south,
                    font=\scriptsize,
                    draw=black!40,
                    fill=white,
                    inner sep=1.5pt,
                    column sep=0.25cm,
                },
            ]
            
            \addplot[solid, color=real_color, line width=0.9pt] table [x=t, y=x_gt] {data/nanodrone_rollout.dat};
            \addlegendentry{Ground Truth}
            \addplot[densely dashed, color=mlp_color, line width=0.80pt] table [x=t, y=x_mlp] {data/nanodrone_rollout.dat};
            \addlegendentry{Res-MLP}
            \addplot[densely dotted, color=lstm_color, line width=0.85pt] table [x=t, y=x_lstm] {data/nanodrone_rollout.dat};
            \addlegendentry{Res-LSTM}
            \addplot[solid, color=ggrnn_color, line width=1.05pt] table [x=t, y=x_ggrnn] {data/nanodrone_rollout.dat};
            \addlegendentry{GGRNN ($A^\star$)}

            \nextgroupplot[
                ylabel={$p_\mathrm{y}$ [m]},
                ymin=-0.8, ymax=0.8,
                ytick={-0.5, 0, 0.5},
            ]
            \addplot[solid, color=real_color, line width=0.9pt] table [x=t, y=y_gt] {data/nanodrone_rollout.dat};
            \addplot[densely dashed, color=mlp_color, line width=0.80pt] table [x=t, y=y_mlp] {data/nanodrone_rollout.dat};
            \addplot[densely dotted, color=lstm_color, line width=0.85pt] table [x=t, y=y_lstm] {data/nanodrone_rollout.dat};
            \addplot[solid, color=ggrnn_color, line width=1.05pt] table [x=t, y=y_ggrnn] {data/nanodrone_rollout.dat};

            \nextgroupplot[
                ylabel={$p_\mathrm{z}$ [m]},
                ymin=0.9, ymax=1.8,
                ytick={1.0, 1.3, 1.6},
            ]
            \addplot[solid, color=real_color, line width=0.9pt] table [x=t, y=z_gt] {data/nanodrone_rollout.dat};
            \addplot[densely dashed, color=mlp_color, line width=0.80pt] table [x=t, y=z_mlp] {data/nanodrone_rollout.dat};
            \addplot[densely dotted, color=lstm_color, line width=0.85pt] table [x=t, y=z_lstm] {data/nanodrone_rollout.dat};
            \addplot[solid, color=ggrnn_color, line width=1.05pt] table [x=t, y=z_ggrnn] {data/nanodrone_rollout.dat};

            \nextgroupplot[
                ylabel={$v_\mathrm{x}$ [m/s]},
                ymin=-2.0, ymax=2.0,
                ytick={-1.5, 0, 1.5},
            ]
            \addplot[solid, color=real_color, line width=0.9pt] table [x=t, y=vx_gt] {data/nanodrone_rollout.dat};
            \addplot[densely dashed, color=mlp_color, line width=0.80pt] table [x=t, y=vx_mlp] {data/nanodrone_rollout.dat};
            \addplot[densely dotted, color=lstm_color, line width=0.85pt] table [x=t, y=vx_lstm] {data/nanodrone_rollout.dat};
            \addplot[solid, color=ggrnn_color, line width=1.05pt] table [x=t, y=vx_ggrnn] {data/nanodrone_rollout.dat};

            \nextgroupplot[
                ylabel={$v_\mathrm{y}$ [m/s]},
                ymin=-2.0, ymax=2.0,
                ytick={-1.5, 0, 1.5},
            ]
            \addplot[solid, color=real_color, line width=0.9pt] table [x=t, y=vy_gt] {data/nanodrone_rollout.dat};
            \addplot[densely dashed, color=mlp_color, line width=0.80pt] table [x=t, y=vy_mlp] {data/nanodrone_rollout.dat};
            \addplot[densely dotted, color=lstm_color, line width=0.85pt] table [x=t, y=vy_lstm] {data/nanodrone_rollout.dat};
            \addplot[solid, color=ggrnn_color, line width=1.05pt] table [x=t, y=vy_ggrnn] {data/nanodrone_rollout.dat};

            \nextgroupplot[
                ylabel={$v_\mathrm{z}$ [m/s]},
                ymin=-0.8, ymax=1.0,
                ytick={-0.5, 0, 0.5},
            ]
            \addplot[solid, color=real_color, line width=0.9pt] table [x=t, y=vz_gt] {data/nanodrone_rollout.dat};
            \addplot[densely dashed, color=mlp_color, line width=0.80pt] table [x=t, y=vz_mlp] {data/nanodrone_rollout.dat};
            \addplot[densely dotted, color=lstm_color, line width=0.85pt] table [x=t, y=vz_lstm] {data/nanodrone_rollout.dat};
            \addplot[solid, color=ggrnn_color, line width=1.05pt] table [x=t, y=vz_ggrnn] {data/nanodrone_rollout.dat};

            \nextgroupplot[
                ylabel={$r_\mathrm{x}$ [rad]},
                ymin=-0.5, ymax=0.5,
                ytick={-0.3, 0, 0.3},
            ]
            \addplot[solid, color=real_color, line width=0.9pt] table [x=t, y=phi_gt] {data/nanodrone_rollout.dat};
            \addplot[densely dashed, color=mlp_color, line width=0.80pt] table [x=t, y=phi_mlp] {data/nanodrone_rollout.dat};
            \addplot[densely dotted, color=lstm_color, line width=0.85pt] table [x=t, y=phi_lstm] {data/nanodrone_rollout.dat};
            \addplot[solid, color=ggrnn_color, line width=1.05pt] table [x=t, y=phi_ggrnn] {data/nanodrone_rollout.dat};

            \nextgroupplot[
                ylabel={$r_\mathrm{y}$ [rad]},
                ymin=-0.5, ymax=0.5,
                ytick={-0.3, 0, 0.3},
            ]
            \addplot[solid, color=real_color, line width=0.9pt] table [x=t, y=theta_gt] {data/nanodrone_rollout.dat};
            \addplot[densely dashed, color=mlp_color, line width=0.80pt] table [x=t, y=theta_mlp] {data/nanodrone_rollout.dat};
            \addplot[densely dotted, color=lstm_color, line width=0.85pt] table [x=t, y=theta_lstm] {data/nanodrone_rollout.dat};
            \addplot[solid, color=ggrnn_color, line width=1.05pt] table [x=t, y=theta_ggrnn] {data/nanodrone_rollout.dat};

            \nextgroupplot[
                ylabel={$r_\mathrm{z}$ [rad]},
                ymin=-0.16, ymax=0.16,
                ytick={-0.1, 0, 0.1},
            ]
            \addplot[solid, color=real_color, line width=0.9pt] table [x=t, y=psi_gt] {data/nanodrone_rollout.dat};
            \addplot[densely dashed, color=mlp_color, line width=0.80pt] table [x=t, y=psi_mlp] {data/nanodrone_rollout.dat};
            \addplot[densely dotted, color=lstm_color, line width=0.85pt] table [x=t, y=psi_lstm] {data/nanodrone_rollout.dat};
            \addplot[solid, color=ggrnn_color, line width=1.05pt] table [x=t, y=psi_ggrnn] {data/nanodrone_rollout.dat};

            \nextgroupplot[
                xlabel={Time [s]},
                ylabel={$\omega_\mathrm{x}$ [rad/s]},
                ymin=-1.3, ymax=1.1,
                ytick={-1.0, 0, 1.0},
            ]
            \addplot[solid, color=real_color, line width=0.9pt] table [x=t, y=wx_gt] {data/nanodrone_rollout.dat};
            \addplot[densely dashed, color=mlp_color, line width=0.80pt] table [x=t, y=wx_mlp] {data/nanodrone_rollout.dat};
            \addplot[densely dotted, color=lstm_color, line width=0.85pt] table [x=t, y=wx_lstm] {data/nanodrone_rollout.dat};
            \addplot[solid, color=ggrnn_color, line width=1.05pt] table [x=t, y=wx_ggrnn] {data/nanodrone_rollout.dat};

            \nextgroupplot[
                xlabel={Time [s]},
                ylabel={$\omega_\mathrm{y}$ [rad/s]},
                ymin=-1.9, ymax=0.9,
                ytick={-1.5, -0.5, 0.5},
            ]
            \addplot[solid, color=real_color, line width=0.9pt] table [x=t, y=wy_gt] {data/nanodrone_rollout.dat};
            \addplot[densely dashed, color=mlp_color, line width=0.80pt] table [x=t, y=wy_mlp] {data/nanodrone_rollout.dat};
            \addplot[densely dotted, color=lstm_color, line width=0.85pt] table [x=t, y=wy_lstm] {data/nanodrone_rollout.dat};
            \addplot[solid, color=ggrnn_color, line width=1.05pt] table [x=t, y=wy_ggrnn] {data/nanodrone_rollout.dat};

            \nextgroupplot[
                xlabel={Time [s]},
                ylabel={$\omega_\mathrm{z}$ [rad/s]},
                ymin=-0.35, ymax=0.10,
                ytick={-0.3, -0.2, -0.1, 0},
            ]
            \addplot[solid, color=real_color, line width=0.9pt] table [x=t, y=wz_gt] {data/nanodrone_rollout.dat};
            \addplot[densely dashed, color=mlp_color, line width=0.80pt] table [x=t, y=wz_mlp] {data/nanodrone_rollout.dat};
            \addplot[densely dotted, color=lstm_color, line width=0.85pt] table [x=t, y=wz_lstm] {data/nanodrone_rollout.dat};
            \addplot[solid, color=ggrnn_color, line width=1.05pt] table [x=t, y=wz_ggrnn] {data/nanodrone_rollout.dat};

        \end{groupplot}
    \end{tikzpicture}
    \vspace{-0.4cm}
    \caption{
    N--DB $50$-step rolling simulations, evaluated at every window origin over a $5\,\mathrm{s}$ segment of the first Melon test flight.
    Rows show position, linear velocity, attitude expressed as roll--pitch--yaw angles, and body-frame angular velocity, respectively.
    A $20$-sample centered rolling mean is applied for visualization, following~\citet{busettoNonlinearSystemIdentification2026}.}
    \vspace{-6pt}
    \label{fig:nanodrone_rollout}
\end{figure*}
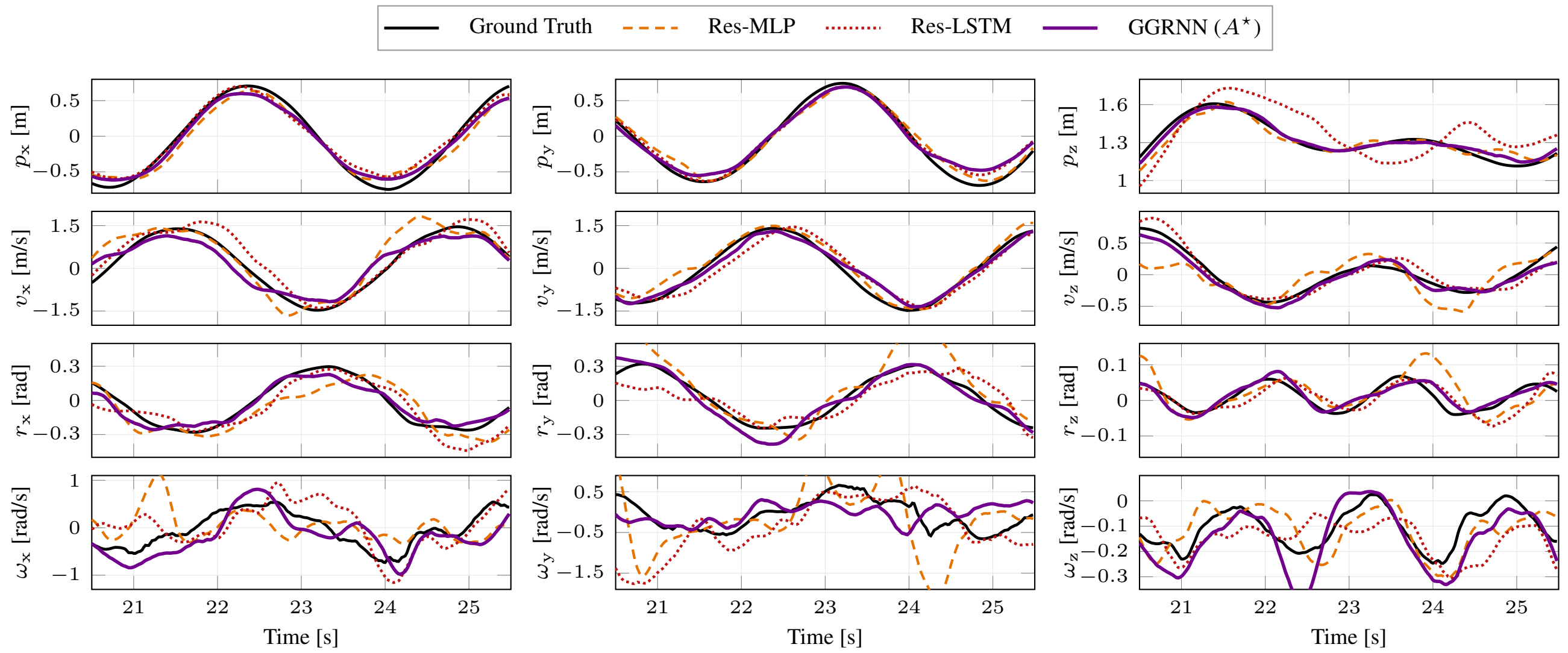

\begin{figure}[ht]
    \vspace*{2pt}
    \centering
    \begin{tikzpicture}
        \begin{groupplot}[
            group style={
                group size=2 by 1,
                horizontal sep=0.30cm,
                ylabels at=edge left,
                yticklabels at=edge left,
            },
            width=0.39\columnwidth,
            height=1.5cm,
            scale only axis,
            boxplot/draw direction=x,
            boxplot/box extend=0.50,
            xmin=-0.04, xmax=1.04,
            ymin=0.4, ymax=4.6,
            xtick={0, 0.25, 0.5, 0.75, 1},
            ytick={1, 2, 3, 4},
            yticklabels={$\bm{\alpha}_\psi^\star$, $\bm{\alpha}_o^\star$, $\bm{\alpha}_r^\star$, $\bm{\alpha}_i^\star$},
            grid=major,
            grid style={dashed, gray!30},
            tick label style={font=\scriptsize},
            label style={font=\scriptsize},
            title style={font=\scriptsize\bfseries, yshift=-4pt},
            xlabel style={yshift=2pt},
        ]
        \nextgroupplot[
            title={W--HB ($n_x = 6$)},
            xlabel={$\alpha$ ($0$: GRU, $1$: LSTM)},
        ]
            \addplot[color=gru_color, densely dashed, line width=0.8pt, forget plot] coordinates {(0,0.4) (0,4.6)};
            \addplot[color=lstm_color, densely dashed, line width=0.8pt, forget plot] coordinates {(1,0.4) (1,4.6)};
            \addplot[
                boxplot={draw position=4},
                color=ggrnn_color,
                fill=ggrnn_color!25,
                line width=0.75pt,
                mark=*,
                mark size=1.0pt,
            ] table[y=wh_alpha_i] {data/alpha_boxplots.dat};
            \addplot[
                boxplot={draw position=3},
                color=ggrnn_color,
                fill=ggrnn_color!25,
                line width=0.75pt,
                mark=*,
                mark size=1.0pt,
            ] table[y=wh_alpha_r] {data/alpha_boxplots.dat};
            \addplot[
                boxplot={draw position=2},
                color=ggrnn_color,
                fill=ggrnn_color!25,
                line width=0.75pt,
                mark=*,
                mark size=1.0pt,
            ] table[y=wh_alpha_o] {data/alpha_boxplots.dat};
            \addplot[
                boxplot={draw position=1},
                color=ggrnn_color,
                fill=ggrnn_color!25,
                line width=0.75pt,
                mark=*,
                mark size=1.0pt,
            ] table[y=wh_alpha_psi] {data/alpha_boxplots.dat};

        \nextgroupplot[
            title={N--DB ($n_x = 64$)},
            xlabel={$\alpha$ ($0$: GRU, $1$: LSTM)},
        ]
            \addplot[color=gru_color, densely dashed, line width=0.8pt, forget plot] coordinates {(0,0.4) (0,4.6)};
            \addplot[color=lstm_color, densely dashed, line width=0.8pt, forget plot] coordinates {(1,0.4) (1,4.6)};
            \addplot[
                boxplot={draw position=4},
                color=ggrnn_color,
                fill=ggrnn_color!25,
                line width=0.75pt,
                mark=*,
                mark size=0.9pt,
            ] table[y=nd_alpha_i] {data/alpha_boxplots.dat};
            \addplot[
                boxplot={draw position=3},
                color=ggrnn_color,
                fill=ggrnn_color!25,
                line width=0.75pt,
                mark=*,
                mark size=0.9pt,
            ] table[y=nd_alpha_r] {data/alpha_boxplots.dat};
            \addplot[
                boxplot={draw position=2},
                color=ggrnn_color,
                fill=ggrnn_color!25,
                line width=0.75pt,
                mark=*,
                mark size=0.9pt,
            ] table[y=nd_alpha_o] {data/alpha_boxplots.dat};
            \addplot[
                boxplot={draw position=1},
                color=ggrnn_color,
                fill=ggrnn_color!25,
                line width=0.75pt,
                mark=*,
                mark size=0.9pt,
            ] table[y=nd_alpha_psi] {data/alpha_boxplots.dat};
        \end{groupplot}
    \end{tikzpicture}
    \vspace{-0.4cm}
    \caption{Distributions of the learned homotopy parameters across recurrent units for the W--HB (left) and the N--DB (right).}
    \label{fig:alpha_boxplots}
\end{figure}
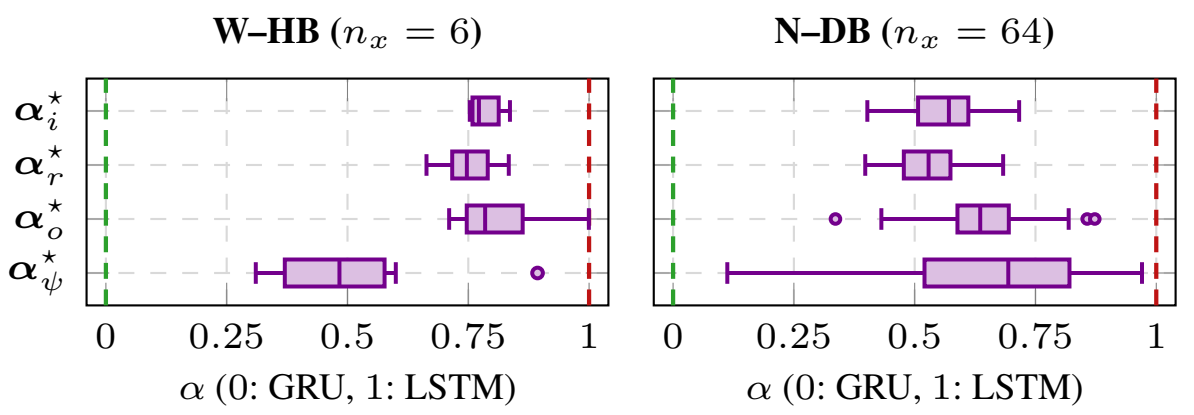

Tables~\ref{tab:results_wh} and~\ref{tab:results_nanodrone} compare the selected GGRNN and its co-trained endpoints with the independently trained literature baselines.

On the W--HB, the co-trained LSTM model improves on both reference networks, while the co-trained GRU model closely matches the reference GRU network.
Stage~2 further improves upon both co-trained endpoints, reducing RMSE by approximately $12\%$ and $21\%$ relative to the independently trained reference GRU and LSTM networks from~\citet{champneysBaselineResultsSelected2024}, respectively.
These results are obtained with a considerably smaller model than the reference networks, both in the state dimension ($6$ against $32$) and in the input ($u_k$ alone against the $32$-dimensional lagged-input vector of Section~\ref{ss:wh_benchmark}).

On the N--DB, the GGRNN improves all four cumulative errors over Res-LSTM and all cumulative errors except the position one over Res-MLP and its co-trained endpoints, achieving the lowest cumulative attitude error $E_r$ overall.
Differently from the W--HB, here the GGRNN and the recurrent reference share the same state dimension $\nx = 64$.
Physics and Hybrid achieve lower cumulative errors for the position $E_p$, velocity $E_v$, and angular velocity $E_\omega$.
In particular, their explicit integration of velocity into position provides a kinematic prior absent from the black-box output maps used here, which most likely is the main cause for the better result for $E_p$.
\figurename{}~\ref{fig:nanodrone_horizon} shows the development of the four errors across the prediction horizon, while \figurename{}~\ref{fig:nanodrone_rollout} illustrates the $50$-step-ahead flight predictions.

%

\figurename{}~\ref{fig:alpha_boxplots} shows the learned homotopy parameters distributions across recurrent units.
The obtained GGRNNs retain mixtures of the endpoint mechanisms: the learned homotopy parameters vary from unit to unit, without a common mixture emerging in either benchmark.
Since these mixtures improve upon at least one endpoint architecture on every output group, the learned homotopies support the premise of this work: blending the GRU and LSTM mechanisms, rather than selecting either one a priori, can improve simulation accuracy.
%
\section{Conclusion}
\label{s:conclusion}

We introduced the GGRNN, an architecture that recovers the GRU and LSTM networks as the exact endpoints of its four elementwise homotopies.
A two-stage training procedure co-trains both endpoints and then optimizes only $4\nx$ homotopy parameters over the frozen network parameters.
For the validation-selected models, this refinement improves simulation accuracy over both endpoints on the W--HB and on three of the four cumulative N--DB errors.
The results support learning a task-dependent combination of recurrent mechanisms rather than choosing either endpoint architecture a priori.
%
\balance
\small
\bibliography{bibliography}

\end{document}